\documentclass[journal]{IEEEtran}
\usepackage{stmaryrd}
\usepackage{amsfonts}
\usepackage{amsmath}
\usepackage{amssymb}
\usepackage{cite}
\usepackage{graphicx}
\usepackage{subfigure}
\usepackage{amsthm}
\usepackage{algorithmic}
\usepackage{algorithm}
\usepackage{stfloats}

\newtheorem{Theorem}{Theorem}
\newtheorem{Lemma}{Lemma}

\begin{document}
%
\title{Synthesis of Stochastic Flow Networks}

\author{Hongchao~Zhou, Ho-Lin Chen,
        and~Jehoshua~Bruck,~\IEEEmembership{Fellow,~IEEE}
\thanks{This work was supported in part by the NSF Expeditions in Computing Program under grant CCF-0832824. This paper was presented in part at IEEE International Symposium on Information Theory (ISIT), Austin, Texas, June 2010. }
\thanks{Hongchao Zhou and Jehoshua Bruck are with the Department
of Electrical Engineering, California Institute of Technology, Pasadena,
CA 91125, e-mail: hzhou@caltech.edu; bruck@caltech.edu.}
\thanks{Ho-lin Chen is with the Department of Electrical Engineering, National Taiwan University, Taipei 106, Taiwan, e-mail: holinc@cc.ee.ntu.edu.tw}
}

\maketitle

\begin{abstract}
A stochastic flow network is a directed graph with incoming edges (inputs) and outgoing edges (outputs), tokens enter through the input edges, travel stochastically in the network, and can exit the network through the output edges. Each node in the network is a splitter, namely, a token can enter a node through an incoming edge and exit on one of the output edges according to a predefined probability distribution.
Stochastic flow networks can be easily implemented by DNA-based chemical reactions, with promising applications in molecular computing and stochastic computing. In this paper, we address a fundamental synthesis question: Given a finite set of possible splitters and an arbitrary rational probability distribution, design a stochastic flow network, such that every token that enters the input edge will exit the outputs with the prescribed probability distribution.

The problem of probability transformation dates back to von Neumann's 1951 work and was followed, among others, by Knuth and Yao in 1976.
Most existing works have been focusing on the ``simulation" of target distributions.  In this paper, we design \emph{optimal-sized} stochastic flow networks for ``synthesizing" target distributions.
It shows that when each splitter has two outgoing edges and is unbiased, an arbitrary rational probability $\frac{a}{b}$ with $a\leq b\leq 2^n$ can be realized by a stochastic flow network of size $n$ that is optimal.
Compared to the other stochastic systems, feedback (cycles in networks) strongly improves the expressibility
of stochastic flow networks.
\end{abstract}


\IEEEpeerreviewmaketitle

\section{Introduction}

The problem of probability transformation dates back to von Neumann \cite{Neumann1951} in 1951, who first considered the problem of simulating an unbiased coin by using a biased coin with unknown probability. He observed that when one focuses on a pair of coin tosses, the events  HT and TH have the same probability (H is for `head' and T is for `tail'); hence, HT produces the output symbol $0$ and TH produces the output symbol $1$. The other two possible events, namely, HH and TT, are ignored, namely, they do not produce any output symbols.
More efficient algorithms for simulating an unbiased coin from a biased coin were proposed by Hoeffding and Simons \cite{Hoeffding1970}, Elias \cite{Elias1972}, Stout and Warren \cite{Warren1984} and Peres \cite{Peres1992}.  In 1976, Knuth and Yao \cite{Knuth1976} presented a simple procedure for generating sequences with arbitrary probability distributions from an unbiased coin (the probability of H and T is $\frac{1}{2}$). They showed that the expected number of coin tosses is upper-bounded by the entropy of the target distribution plus two.
 Han and Hoshi \cite{Han1997} and Abrahams \cite{Abrahams1996} generalized their approach and demonstrated how to generate an arbitrary probability distribution using a general $M$-sided biased coin.
All these works have been focusing on the ``simulation" side of probability transformation, and  their goal is to minimize
the expected number of coin tosses for generating a certain number of target distributions.

\begin{figure}[!b]
\centering
\includegraphics[height=1.6in]{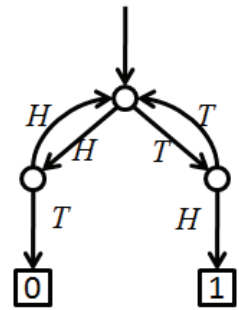}
\caption{An instance of stochastic flow network that consists of three $p$-splitters for any $p$
and generates probability $\frac{1}{2}$.}
\label{fig_example1}
\end{figure}


There are a few works that considered the problem of probability transformation from a synthetic perspective, namely, designing a physical system
for ``synthesizing" target distributions, by connecting certain probabilistic elements. Such probabilistic elements
can be electrical ones based on internal thermal noise or molecular ones based on inherent randomness in chemical reactions.
In this scenario, the size of the construction becomes a central issue.
In 1962,
Gill \cite{Gill62}\cite{Gill63} discussed
the problem of generating rational probabilities using a sequential
state machine. Later, Sheng \cite{Sheng1965} considered applying threshold logic elements as a discrete probability transformer.
Recently, Wilhelm and Bruck \cite{Wilhelm2008} proposed a procedure for synthesizing stochastic switching circuits to realize desired discrete probabilities. More properties and constructions of stochastic switching circuits were studied by Zhou, Loh and Bruck \cite{Zhou09,Loh2009,Zhou2011};
Qian et. al. \cite{Qian2011} studied combinational logic for transforming a set of given probabilities into target probabilities.
Motivated by stochastic computing based on chemical reaction networks\cite{Soloveichik2009}, in this paper we study stochastic flow networks. A stochastic flow network is a directed graph with incoming edges (inputs) and outgoing edges (outputs), tokens enter through the input edges, travel stochastically in the network and can exit the network through the output edges. Each node in the network is a splitter, namely, a token can enter a node through an incoming edge and exit on one of the output edges according to a predefined probability distribution. We address a fundamental synthesis question: Given a finite set of possible splitters and an arbitrary rational probability distribution, design an \emph{optimal-sized} stochastic flow network, such that every token that enters the input edge will exit the outputs with the prescribed probability distribution.

Stochastic flow networks can be easily implemented by chemical reaction networks, where each splitter corresponds to two types of molecules, and incoming tokens (another type of molecules) can react with both, hence react with one of them with a certain probability.
Compared to the synthetic stochastic systems described above, stochastic flow networks demonstrate strong powers in expressing an arbitrary rational target distribution.
Fig. \ref{fig_example1} depicts von Neumann's algorithm in the language a stochastic flow network that consists of three $p$-splitters for any $p$ and generates probability $\frac{1}{2}$.
Here, a $p$-splitter indicates a splitter with two outgoing edges with probabilities $p$ and $(1-p)$.
In this construction, we have two outputs  $\{\beta_1,\beta_2\}=\{0,1\}$ (corresponding to the labels $0$ and $1$, respectively).
For each incoming token, it has the same probability $pq$ to reach either output $0$ or output $1$ directly, and it has probability $1-2pq$
to come back to the starting point. Eventually, the probability for the token to reach each of the outputs is $\frac{1}{2}$.
In general, the outputs of  a stochastic flow network have labels denoted by $\{\beta_1,\beta_2,...,\beta_m\}$. A token will reach an output $\beta_k$ $(1\leq k\leq m)$ with probability $q_k$, and we call $q_k$ the probability of $\beta_k$ and call $\{q_1,q_2,...,q_m\}$ the output probability distribution of the network, where $\sum_{k=1}^m q_k=1$.

In this paper we assume, without loss of generality, that the probability of each splitter is $\frac{1}{2}$ ($\frac{1}{2}$-splitters can be implemented using three $p$-splitters for any $p$). Our goal is to realize the target probabilities or distributions by constructing a network of minimum size. In addition, we study the expected latency, namely the expected number of splitters a token need to pass before reaching the output (or we call it the expected operating time).

The main contributions of the paper are
\begin{enumerate}
\item \emph{General optimal construction:} For any desired rational probability, an \emph{optimal-sized} construction of stochastic flow network is provided.
\item \emph{The power of feedback:} We show that with feedback (loops),
 stochastic flow networks can generate significantly more probabilities than those without feedback.
\item \emph{Constructions with well-bounded expected latency:} We give two constructions whose expected latencies are well-bounded by constants.
As a price, they use a few more splitters than the optimal-sized one.
\item \emph{Constructions for arbitrary rational distributions: } We generalize our constructions so that they can realize an arbitrary rational probability distribution $\{q_1,q_2,...,q_m\}$.
\end{enumerate}

The remainder of this paper is organized as follows. In Section \ref{section_preliminary} we introduce some preliminaries including Knuth and Yao's scheme and a few mathematical tools
for calculating the distribution of a given stochastic flow network. Section \ref{section_part1} introduces an optimal-sized construction of stochastic flow networks for synthesizing an arbitrary rational probability, and it demonstrates that feedback significantly enhances the expressibility of stochastic flow networks.
Section \ref{section_properties} analyzes the expected latency of the optimal-sized construction.
Section \ref{section_upg} gives two constructions whose expected latencies are upper bounded by constants.
Section \ref{section_distributions} presents the generalizations of our results to arbitrary rational probability distributions.
The concluding remarks and the comparison of different stochastic systems are given in Section \ref{section_conclusion}.

\section{Preliminaries}
\label{section_preliminary}

In this section, we introduce some preliminaries, including Knuth and Yao's scheme for simulating an arbitrary distribution from a biased coin, and how using absorbing Markov chains or Mason' Rule to calculate the output distribution of a given stochastic flow network.

\subsection{Knuth and Yao's Scheme}

In 1976, Knuth and Yao proposed a simple procedure for simulating an arbitrary distribution from an unbiased coin (the probability of H and T is $\frac{1}{2}$) \cite{Knuth1976}. They
introduced a concept called generating tree for representing the algorithm \cite{Cover2006}. The leaves of the tree are marked by the output symbols, and the path from the root node to the leaves
indicates the sequences of bits generated by the unbiased coin. Starting from the root node, the scheme selects edges to follow based on
the coin tosses until it reaches one of the leaves. Then it outputs the symbol marked on that leaf.

\begin{figure}[!b]
\centering
\includegraphics[height=1.8in]{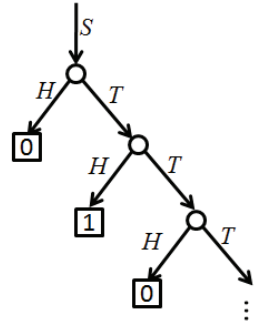}
\caption{The generating tree to generate a $(\frac{2}{3},\frac{1}{3})$ distribution.}
\label{fig_example2}
\end{figure}

In general, we assume that the target distribution is $\{p_1,p_2,..., p_m\}$.
Since all the leaves of the tree have probabilities of the form $2^{-k}$ (if the depth of the leaf is $k$), we split each
probability $p_i$ into atoms of this form. Specifically,
let the binary expansion of the probability $p_i$ be
$$p_i=\sum_{j\geq 1} p_i^{(j)},$$
where $p_i^{(j)}=2^{-j}$ or $0$. Then for each probability $p_i$, we get a group of atoms
$\{p_i^{(j)}:j\geq 1\}$. For these atoms, we allot them to leaves with label $\beta_i$
on the tree. Hence, the probability of generating $\beta_i$ is $p_i$.
We can see that the depths of all the atoms satisfy the Kraft inequality \cite{Cover2006}, i.e.,
$$\sum_{i=1}^m \sum_{j\geq 1} p_i^{(j)}=1.$$
So we can always construct such a tree with all the atoms allotted. Knuth and Yao showed that the expected number of fair
bits required by the procedure (i.e. the expected depth of the tree) to generate a random
variable $X$ with distribution $\{p_1, p_2, ..., p_m\}$ lies between $H(X)$
and $H(X)+2$ where $H(X)$ is the entropy of the target distribution.

Fig. \ref{fig_example2} depicts a generating tree that generates a distribution $\{\frac{2}{3},\frac{1}{3}\}$, where
the atoms for $\frac{2}{3}$ are $\{\frac{1}{2},\frac{1}{8},\frac{1}{32},...\}$, and
the atoms for $\frac{1}{3}$ are $\{\frac{1}{4},\frac{1}{16},\frac{1}{64},...\}$.  We see that the construction of generating trees is, in some sense,
a special case of stochastic flow networks that without cycles. If we consider each node in the generating tree as a splitter, then
each token that enters the tree from the root node will reach the outputs with the target distribution.
While Knuth and Yao's scheme aims to minimize
the expected depth of the tree (or in our framework, we call it the expected latency of the network), our goal is to optimize the size of the construction, i.e., the number of nodes in the network.

\subsection{Absorbing Markov Chain}

Let's consider a stochastic flow network with $n$ splitters and $m$ outputs, in which each splitter is associated with a state number in $\{1,2,...,n\}$ and each output is associated with a
state number in $\{n+1,n+2,...,n+m\}$. When a token reaches splitter $i$ with $1\leq i\leq n$, we say that the current state of this network is $i$. When it reaches output $k$ with $1\leq k\leq m$, we say that
the current state of this network is $n+k$. Note that the current state of the network only depends on the last state, and when the token reach one output it will stay there forever. So we can describe token flows in this network using an absorbing Markov chain. If the current state of the network is $i$, then the probability of reaching state $j$ at the next instant of time is given by $p_{ij}$. Here, $p_{ij}=p_H$ ($p_{ij}=p_T$) if and only if state $i$ and state $j$ is connected by an edge $H$ ($T$).

Clearly, the network with $n$ splitters and $m$ outputs with different labels can be described by an absorbing Markov chain, where the first $n$ states are transient states and the last $m$ states are absorbing states. And we have
$$\begin{array}{rcl}
    \sum_{j=1}^{n+m}p_{ij}=1 &\quad & i=1,2,...,n+m, \\
    p_{ij}=0 &\quad& \forall i>n \textrm{ and } i\neq j, \\
    p_{ii}=1 &\quad&  \forall i>n.
  \end{array}
$$

The transition matrix of this Markov chain is given by
$$P=
\begin{array}{cc}
& n \quad m \\
\begin{array}{c}
      n \\
      m
    \end{array} & \left(
      \begin{array}{cc}
      Q & R \\
      0 & I \\
      \end{array}
    \right)
\end{array}
$$
where $Q$ is an $n\times n$ matrix, $R$ is an $n\times m$ matrix, $0$ is an $m\times n$ zeros matrix and $I$ is an $m\times m$ identity matrix.

Let $B_{ij}$ be the probability for an absorbing Markov chain reaching the state $j+n$ if
it starts in the transient state $i$. Then $B$ is an $n\times m$ matrix, and
$$B=(I-Q)^{-1}R.$$

Assume this Markov chain starts from state $1$ and let $S_j$ be the probability for it reaching the absorbing state $j+n$. Then $S$ is the distribution of the network
$$S=[1,0,...,0]B= e_1(I-Q)^{-1}R.$$

\begin{figure}[!t]
\centering
\includegraphics[height=1.6in]{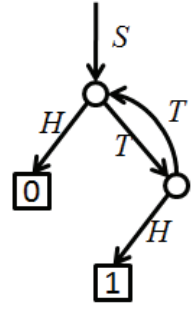}
\caption{The stochastic flow network to generate a $(\frac{2}{3},\frac{1}{3})$ distribution.}
\label{fig_example3}
\end{figure}

Given a stochastic flow network, we can use the formula above to calculate its probability distribution. For example, the transition matrix of the network in Fig. \ref{fig_example3} is

$$P=
\left(
  \begin{array}{cccc}
0 & \frac{1}{2} & \frac{1}{2} & 0\\
\frac{1}{2} & 0 & 0 & \frac{1}{2}\\
0 & 0 & 1 & 0\\
0 & 0 & 0 & 1\\
  \end{array}
\right)
$$

From which we can obtain the probability distribution
$$S= e_1(I-Q)^{-1}R= \left(
      \begin{array}{cc}
        \frac{2}{3} & \frac{1}{3} \\
      \end{array}
    \right).
$$

\subsection{Mason's Rule}

Mason's gain rule is a method used in control theory to find the transfer function of a given control system. It can be applied to any signal flow graph. Generally,  we describe it as follows (see more details about Mason's rule in \cite{Valkenburg1974}):

Let $H(z)$ denote the transfer function of a signal flow graph. Define the following notations:
\begin{enumerate}
  \item $\Delta(z)=$ determinant of the graph.
  \item $L=$ number of forward paths, with $P_k(z)$, $1\leq k\leq L$ denoting the forward path gains.
  \item $\Delta_k(z)=$ determinant of the graph that remains after deleting the $k$th forward path $P_k(z)$.
\end{enumerate}

To calculate the determinant of a graph $\Delta(z)$, we list all the loops in the graph and their gains denoted by $L_i$, all pairs of non-touching loops
$L_iL_j$, all pairwise non-touching loops $L_iL_jL_k$, and so forth. Then
$$\Delta(z)=1-\sum_{i: \textrm{loops}} L_i +\sum_{(i,j): \textrm{non-touching}} L_iL_j -...$$

The transfer function is
$$H(z)=\frac{\sum_{k=1}^L P_k(z)\Delta_k(z)}{\Delta(z)}, $$
called Mason's rule.

Let's treat a stochastic flow network as a control system with input $U(z)=1$. Applying Mason's rule to this system, we can get the probability that one token reaches output $k$ with $1\leq k\leq m$. Also having the network in Fig. \ref{fig_example3} as an example:  In this network, we want to calculate the probability for a token to reach output $1$ (for short, we call it as the probability of $1$). Since there is only one loop with gain $=\frac{1}{4}$ and only one forward path with forward gain $\frac{1}{4}$, we can obtain that the probability of $1$ is
$$P=\frac{\frac{1}{4}}{1-\frac{1}{4}}=\frac{1}{3},$$
which accords with the result of absorbing Markov chains. In fact, it can be proved that the Mason's rule
and the matrix form based on absorbing Markov chains are equivalent.

\section{Optimal-Sized Construction and Feedback}
\label{section_part1}

In this section we present an optimal-sized construction of stochastic flow networks. It consists of splitters with probability 1/2 and computes an arbitrary rational probability.
We demonstrate that feedback (loops) in stochastic flow networks significantly enhance their expressibility. To see that,
let's first study stochastic flow networks without loops, and then those with loops.

\subsection{Loop-free networks}

Here, we want to study the expressive power of loop-free networks.  We say that there are no loops in a network if no tokens
can pass any position in the network more than once. For loop-free networks, we have the following theorem:
\begin{Theorem}
For a loop-free network with $n$ $\frac{1}{2}$-splitters, any probability
$\frac{x}{2^n}$ with integer $x (0\leq x\leq 2^n)$ can be realized, and only probabilities
$\frac{x}{2^n}$ with integer $x (0\leq x\leq 2^n)$ can be realized.
\label{theorem_loopfree}
\end{Theorem}

\proof a) In order to prove that all probability $\frac{x}{2^n}$ with integer $x (0\leq x\leq 2^n)$ can be realized,
we only need to provide the constructions of the networks.

\begin{figure}[!t]
\centering
\includegraphics[width=1.6in]{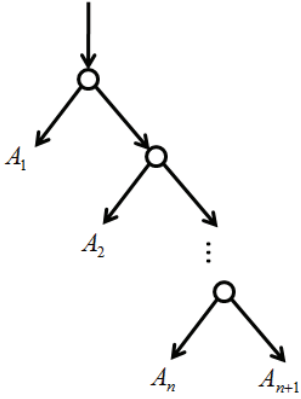}
\caption{Tree structure used to realize probability $\frac{x}{2^n}$ for an integer $x (0\leq x\leq 2^n)$ .}
\label{fig_tree1}
\end{figure}

\begin{enumerate}
  \item Construct a tree, as shown in Fig. \ref{fig_tree1}. In this tree structure, each token will
  reach $A_i (1\leq i\leq n)$ with probability $2^{-i}$, and reach $A_{n+1}$ with probability
$2^{-n}$.
  \item Let
  $\frac{x}{2^n}=\sum_{i=1}^{n} \gamma_i 2^{-i}$,
 where $\gamma_i=0$ or $1$. For each $j$ with $1\leq j\leq n$, $\gamma_j=1$, we connect $A_j$ to  output $0$; otherwise, we connect $A_j$ to output $1$.
 Then we connect $A_{n+1}$ to output $1$. Eventually, the probability for a token to reach output $0$ is
 $$P=\sum_{j=1}^n\frac{\gamma_{n-j}}{2^j}= \sum_{i=0}^{n-1} \frac{\gamma_i}{2^{n-i}}=\frac{x}{2^n}.$$
 \end{enumerate}
Using the procedure above, we can construct a network such that its probability is $\frac{x}{2^n}$. Actually,
it is a special case of Knuth and Yao's construction \cite{Knuth1976}.

b) Now, we prove that only probability
$\frac{x}{2^n}$ with integer $x (0\leq x\leq 2^n)$ can be realized. If this is true, then
$\frac{x}{2^n}$ with odd $x$ cannot be realized with less than $n$ splitters. It means that
in the construction above, the network size $n$ is optimal.

According to Mason's rule, for a network without loops,
the probability for a token reaching one output is
$$P=\sum_k P_k,$$
where $P_k$ is the path gain of a forward path from the root to the output. Given $n$ splitters,  the length of each forward path
should be at most $n$. Otherwise,  there must be a loop along this forward path (have to pass the same
splitter for at least two times). For each $k$, $P_k$ can be written as $\frac{x_k}{2^n}$ for some $x_k$. As a result,
we can get that $P$ can be written as $\frac{x}{2^n}$ for some $x$.
\hfill\QED

\subsection{Networks with loops}

We showed that stochastic flow networks without loops can only realize binary probabilities. Here, we show that
feedback (loops) plays an important rule in enhancing their expressibility. For example, with feedback, we can realize probability $\frac{2}{3}$ with only two splitters, as shown in
Fig. \ref{fig_example3}. But without loops, it is impossible (or requires an infinite number of splitters) to
realize $\frac{2}{3}$. More generally,  for any desired rational probability $\frac{a}{b}$ with integers $0\leq a\leq b\leq 2^n$,  we have the following theorem:

\begin{Theorem}
For a network with $n$ $\frac{1}{2}$-splitters, any rational probability $\frac{a}{b}$ with integers $0\leq a\leq b\leq 2^n$ can be realized
, and only rational probabilities $\frac{a}{b}$ with integers $0\leq a\leq b\leq 2^n$ can be realized.
\label{theorem_gereral1}
\end{Theorem}

\proof a) We prove that all rational probability $\frac{a}{b}$ with integers $0\leq a\leq b\leq 2^n$ can be realized. When
$b=2^n$, the problem becomes trivial due to the result of Theorem \ref{theorem_loopfree}. In the following proof, without loss of generality (w.l.o.g),
we only consider the case in which $2^{n-1}<b<2^n$ for some $n$.

We first show that all probability distributions $\{\frac{x}{2^n},\frac{y}{2^n},\frac{z}{2^n}\}$ with integers $x,y,z$ s.t. $(x+y+z=2^n)$
can be realized with $n$ splitters. Now let's construct the network iteratively.

When $n=1$, by enumerating all the possible connections, we can verify that all the following probability distributions can be realized:
$$\{0,0,1\},\{0,1,0\},\{1,0,0\},$$
$$\{0,\frac{1}{2},\frac{1}{2}\},\{\frac{1}{2},0,\frac{1}{2}\},\{\frac{1}{2},\frac{1}{2},0\}.$$
So all the probability distributions $\{\frac{x}{2},\frac{y}{2},\frac{z}{2}\}$ with integers $x,y,z$ s.t. $(x+y+z=2)$
can be realized.

Assume that all the probability distribution $\{\frac{x}{2^k},\frac{y}{2^k},\frac{z}{2^k}\}$ with integers $x,y,z$ s.t. $(x+y+z=2^k)$ can be realized by a network with $k$ splitters, then we show that
any desired probability distribution $\{\frac{x}{2^{k+1}},\frac{y}{2^{k+1}},\frac{z}{2^{k+1}}\}$
s.t. $x+y+z=2^{k+1}$ can be realized with one more splitter.
Since $x+y+z=2^{k+1}$, at least one of $x,y,z$ is even. W.l.o.g, we let $x$ be even. Then there are two cases to consider: either both $y$ and $z$ are even, or both $y$ and $z$ are odd.

When both $y$ and $z$ are even, the problem is trivial since the desired probability distribution can be written as
$\{\frac{x/2}{2^k},\frac{y/2}{2^k},\frac{z/2}{2^k}\}$,
which can be realized by a network with $k$ splitters.

\begin{figure}[!t]
\centerline{\subfigure[]{\includegraphics[width=2.2in]{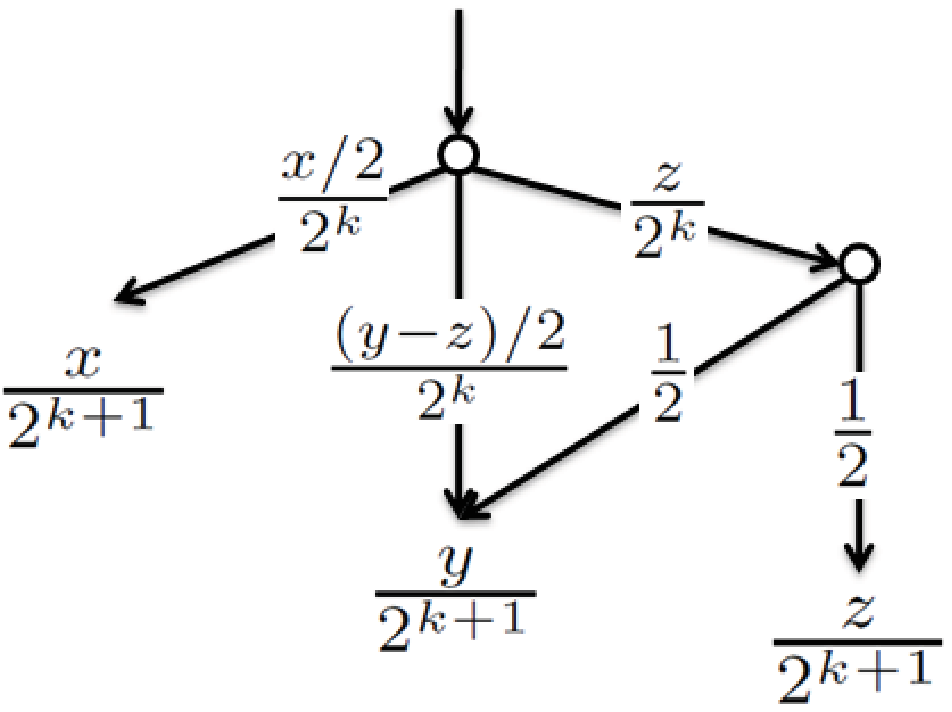} \label{fig_loop3}}}
\centerline{\subfigure[]{\includegraphics[width=2in]{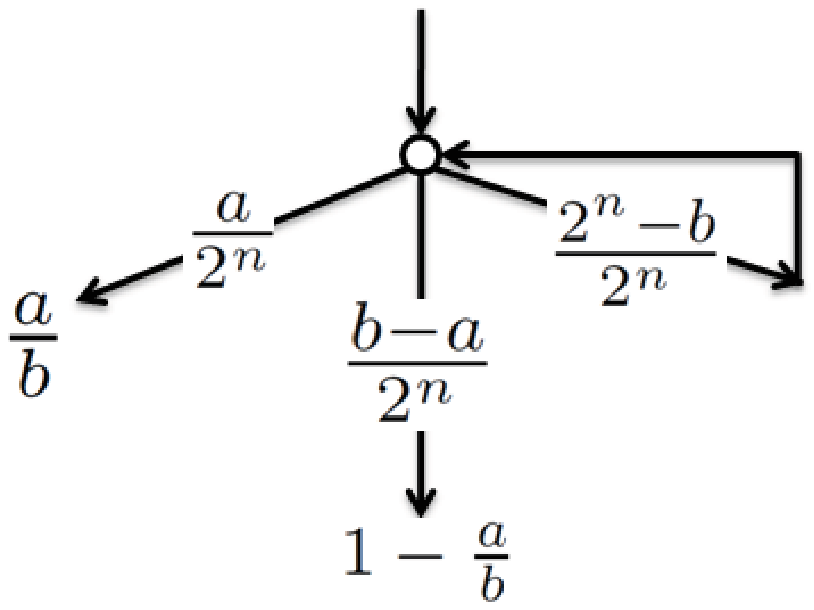} \label{fig_loop6}}}
\caption{(a) The network to realize  $\{\frac{x}{2^{k+1}},\frac{y}{2^{k+1}},\frac{z}{2^{k+1}}\}$ iteratively. (b) The network to realize  $\{\frac{a}{b},1-\frac{a}{b}\}$. } \label{fig_loopssss}
\end{figure}

When both $y$ and $z$ are odd, w.l.o.g, we assume that $z\leq y$. In this case,  we construct a network to realize probability distribution
$\{\frac{x/2}{2^k}, \frac{(y-z)/2}{2^k}, \frac{z}{2^k}\}$ with $k$ splitters.
By connecting the last output with probability $\frac{z}{2^k}$ to an additional splitter, we can get
a new distribution $\{\frac{x/2}{2^k}, \frac{(y-z)/2}{2^k}, \frac{z}{2^{k+1}},\frac{z}{2^{k+1}}\}$.
If we consider the second and the third output as a single output, then
 we can get a new network in
Fig. \ref{fig_loop3}, whose probability distribution is $\{\frac{x}{2^{k+1}},\frac{y}{2^{k+1}},\frac{z}{2^{k+1}}\}$.

Hence, for any probability distribution
$\{\frac{x}{2^n},\frac{y}{2^n},\frac{z}{2^n}\}$ with $x+y+z=2^n$, we can always construct a
network with $n$ splitters to realize it.

Now, in order to realize probability $\frac{a}{b}$ with $2^{n-1}<b<2^n$ for some $n$, we can construct a network with probability distribution
$\{\frac{a}{2^n},\frac{b-a}{2^n},\frac{2^n-b}{2^n}\}$ with $n$ splitters and connect the last output (output $2$) to the starting
point of the network, as shown in Fig. \ref{fig_loop6}. Using the method of absorbing Markov chains, we can obtain that the probability for a token to reach output $0$ is $\frac{a}{b}$. A simple understanding for this result is that: (1) the ratio of the probabilities for a token to reach the first output and the second output is $\frac{a}{2^n}:\frac{b-a}{2^n}$ that equals
$a:(b-a)$  (2) the sum of these two probabilities is $1$, since the tokens will finally reach one of the two outputs.

b) Now we prove that with $n$ splitters, only rational probability $\frac{a}{b}$ with integers $0\leq a\leq b\leq 2^n$ can be realized.
For any flow network with $n$ splitters, it can be described as an absorbing Markov chain with $n$ transient states and
$2$ absorbing states, whose transition matrix $P$ can be written as
$$P=\left(
      \begin{array}{ccccc}
        p_{11} &  \ldots &p_{1n}& p_{1(n+1)} & p_{1(n+2)} \\
         \vdots & \ddots &\vdots& \vdots  & \vdots\\
        p_{n1} & \ldots & p_{nn} & p_{n(n+1)} & p_{n(n+2)}\\
        0 & \ldots & 0 & 1 & 0 \\
        0 & \ldots & 0 & 0 & 1 \\
      \end{array}
    \right)
$$
where each row consists of two $\frac{1}{2}$ entries and $n$ zeros.

Let
$$Q=\left(
      \begin{array}{ccc}
        p_{11} &  \ldots &p_{1n} \\
         \vdots & \ddots &\vdots\\
        p_{n1} & \ldots & p_{nn} \\
      \end{array}
    \right), R=\left(
      \begin{array}{cc}
         p_{1(n+1)} & p_{1(n+2)} \\
          \vdots  & \vdots\\
         p_{n(n+1)} & p_{n(n+2)}\\
      \end{array}
    \right)$$
then the probability distribution of the network can be written as
$$e_1(I-Q)^{-1} R.$$

In order to prove the result in the theorem, we only need to prove that $(I-Q)^{-1}R$ can be written as
$\frac{1}{b}A$ with $b\leq 2^n$, where $A$ is an integer matrix (all the entries in $A$ are integers).

Let $K=I-Q$, we know that $K$ is invertible if and only $det(K)\neq 0$. In this case, we have
$$(K^{-1})_{ij}= \frac{K_{ji}}{det(K)},$$
where $K_{ji}$ is defined as the determinant of the square matrix of order $(n-1)$ obtained from $K$
by removing the $i^{th}$ row and the $j^{th}$ column multiplied by $(-1)^{i+j}$.

Since each entry of $K$ is chosen from $\{0,\frac{1}{2},1\}$, $K_{ji}$ can be written as $\frac{k_{ji}}{2^{n-1}}$
for some integer $k_{ji}$ and $det(K)$ can be written as $\frac{b}{2^n}$ for some integer $b$. According to Lemma \ref{Lemma_appendix1} in the appendix,
we have $0\leq det(K)\leq 1$, which leads us to $0<b\leq 2^n$ (note that $det(K)\neq 0$).

Then, we have that
\begin{eqnarray*}
K^{-1}&=&\frac{1}{det(K)} \left(
                            \begin{array}{cccc}
                              K_{11} & K_{21} & \ldots & K_{n1} \\
                              K_{12} & K_{22} & \ldots & K_{n2} \\
                              \vdots & \vdots & \ddots & \vdots \\
                              K_{1n} & K_{2n} & \ldots & K_{nn} \\
                            \end{array}
                          \right)\\
&=& \frac{2}{b} \left(
                            \begin{array}{cccc}
                              k_{11} & k_{21} & \ldots & k_{n1} \\
                              k_{12} & k_{22} & \ldots & k_{n2} \\
                              \vdots & \vdots & \ddots & \vdots \\
                              k_{1n} & k_{2n} & \ldots & k_{nn} \\
                            \end{array}
                          \right)
\end{eqnarray*}

Since each entry of $R$ is also in $\{0,\frac{1}{2},1\}$, we know that
$$2R=\left(
       \begin{array}{cc}
         r_{11} & r_{12} \\
         r_{21} & r_{22} \\
         \vdots & \vdots \\
         r_{n1} & r_{n2} \\
       \end{array}
     \right)
$$ is an integer matrix.

As a result
\begin{eqnarray*}
K^{-1}R
&=& \frac{2R}{b} \left(
                            \begin{array}{cccc}
                              k_{11} & k_{21} & \ldots & k_{n1} \\
                              k_{12} & k_{22} & \ldots & k_{n2} \\
                              \vdots & \vdots & \ddots & \vdots \\
                              k_{1n} & k_{2n} & \ldots & k_{nn} \\
                            \end{array}
                          \right)\\
&=& \frac{1}{b}\left(
                            \begin{array}{cccc}
                              k_{11} & k_{21} & \ldots & k_{n1} \\
                              k_{12} & k_{22} & \ldots & k_{n2} \\
                              \vdots & \vdots & \ddots & \vdots \\
                              k_{1n} & k_{2n} & \ldots & k_{nn} \\
                            \end{array}
                          \right)\left(
       \begin{array}{cc}
         r_{11} & r_{12} \\
         r_{21} & r_{22} \\
         \vdots & \vdots \\
         r_{n1} & r_{n2} \\
       \end{array}
     \right)\\
&=& \frac{A}{b},
\end{eqnarray*}
where each entry of $A$ is an integer. So all the probabilities in the final distribution are of the form $\frac{a}{b}$.

This completes the proof.
\hfill\QED

Based on the method in the theorem above, we can realize any arbitrary rational probability with an optimal-sized network. The construction has two steps:
\begin{enumerate}
  \item Construct a network with output distribution $\{\frac{a}{2^n},\frac{b-a}{2^n},\frac{2^n-b}{2^n}\}$ iteratively
  using at most $n$ splitters.
  \item Connect the last output to the starting point, such that the distribution of the resulting network is $\{\frac{a}{b},\frac{b-a}{b}\}$.
\end{enumerate}

\begin{figure}[!t]
\centerline{\subfigure[]{\includegraphics[width=1.5in]{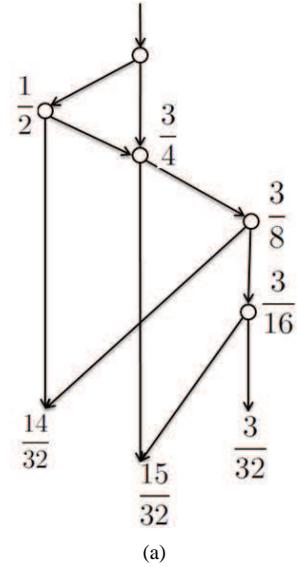} \label{fig_loop5}}}
\centerline{\subfigure[]{\includegraphics[width=1.8in]{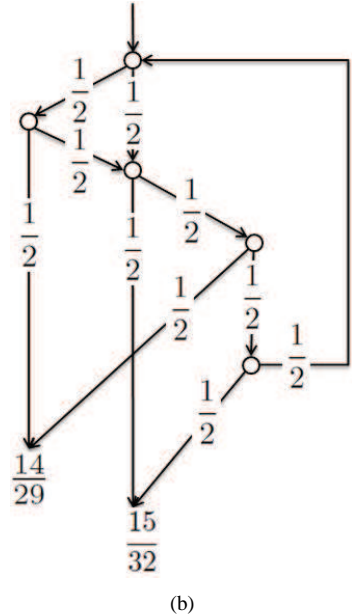} \label{fig_loop7}}}
\caption{(a) The network to realize probability distribution $\{\frac{14}{32},\frac{15}{32},\frac{3}{32}\}$ (b)The network to realize probability $\frac{14}{29}$. } \label{fig_loops}
\end{figure}

When $b=2^n$ for some $n$, the construction above is exactly the generating tree construction
in the Knuth and Yao's scheme as described in Section \ref{section_preliminary}.  Now, assume we want to realize
probability $\frac{14}{29}$. We can first generate a probability distribution $\{\frac{14}{32},\frac{15}{32},\frac{3}{32}\}$, which can be realized by adding one splitter
to a network with probability distribution $\{\frac{7}{16},\frac{6}{16},\frac{3}{16}\}$... Recursively, we can have the following probability distributions:
$$\{\frac{14}{32},\frac{15}{32},\frac{3}{32}\}\rightarrow  \{\frac{7}{16},\frac{6}{16},\frac{3}{16}\} \rightarrow \{\frac{2}{8}, \frac{3}{8}, \frac{3}{8}\}$$
$$\rightarrow \{\frac{1}{4},0,\frac{3}{4}\} \rightarrow \{\frac{1}{2}, 0, \frac{1}{2}\}.$$

As a result, we get a network to generate probability distribution $\{\frac{14}{32},\frac{15}{32},\frac{3}{32}\}$, as shown in Fig. \ref{fig_loop5}, where only $5$ splitters are used. Connecting the last output to the starting point results in the network in Fig. \ref{fig_loop7} with probability $\frac{14}{29}$. Comparing the results in Theorem \ref{theorem_gereral1} with
those in Theorem \ref{theorem_loopfree}, we see that introducing
loops into networks can strongly enhance their expressibility.

\section{Expected latency of Optimal Construction}
\label{section_properties}

Besides of network size, anther important issue of a stochastic flow network is the expected operating time, or we call it
expected latency, defined as the expected number of splitters a token need to pass
before reaching one of the outputs. For the optimal-sized construction proposed in the above section, we have the following results about its expected latency.

\begin{Theorem}
Given a network with rational probability $\frac{a}{b}$ with $b\leq 2^n$ constructed using the optimal-sized construction,
its expected latency $ET$ is upper bounded by \footnote{By making the construction more sophisticated, we can reduce the upper bound to $(\frac{n}{2}+\frac{3}{4})\frac{2^n}{b}$.}$$ET\leq (\frac{3n}{4}+\frac{1}{4})\frac{2^n}{b}<\frac{3n}{2}+\frac{1}{2}.$$
\end{Theorem}

\proof
For the optimal-sized construction, we first prove that the expected latency of the network with distribution $\{\frac{a}{2^n},\frac{b-a}{2^n},\frac{2^n-b}{2^n}\}$ is bounded by
$\frac{3n}{4}+\frac{1}{4}$.

Let's prove this by induction. When $n=0$ or $n=1$, it is easy to see that this conclusion is true. Assume when $n=k$, this conclusion is true, we want to show that
the conclusion still holds for $n=k+2$. Note that in the optimal-sized construction, a network with size $k+2$ can be constructed by adding two more splitters to a network with size $k$.
Let $T_k$ denote the latency of the network with size $k$, then
$$E[T_{k+2}]=E[T_k]+p_1+p_2,$$
where $p_1$ is the probability for a token to reach the first additional splitter and $p_2$ is the probability for a token to reach
the second additional splitter. Assume the distribution of the network with size $k$ is $\{q_1,q_2,q_3\}$, then
$$p_1+p_2\leq \max_{i\neq j}(q_i+(\frac{q_i}{2}+q_j))\leq \frac{3}{2}.$$
So the conclusion is true for $n=k+2$. By induction, we know that it holds for all $n\in \{0,1,2,...\}$.

Secondly, we prove that if the expected latency of the network with distribution $\{q_1,q_2,q_3\}$ is $ET'$, then by connecting its last output to its starting point, we can
  get a network such that its expected latency is $ET=\frac{ET'}{q_1+q_2}$. This conclusion can be obtained immediately from
$$ET=ET'+q_3(ET).$$

This completes the proof.
\hfill\QED

\begin{Theorem}
There exists a network of size $n$ constructed using the optimal-sized construction such that its expected latency $ET$ is lower bounded by
$$ET\geq \frac{n}{3}+\frac{2}{3}.$$
\end{Theorem}

\begin{figure}[!t]
\centering
\includegraphics[width=1.4in]{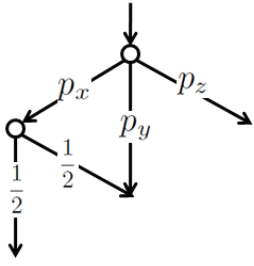}
\caption{Illustration for the construction of a network with unbounded expected latency. Here, we have  $p_x\geq p_y\geq p_z$.}
\label{fig_unBound1}
\end{figure}

\proof We only need to construct a network with distribution $\{\frac{x}{2^n}, \frac{y}{2^n}, \frac{z}{2^n}\}$ for some integers $x,y,z$ such that
its expected latency is lower bounded by $\frac{n}{3}+\frac{2}{3}$.

\begin{table*}
  \centering
  \renewcommand{\arraystretch}{2}
  \begin{tabular}{|c|c|c|c|}
    \hline
     & Optimal-Sized Construction &  Size-Relaxed Construction & Latency-Oriented Construction \\
     \hline
    Network size     & $\leq n$ & $\leq n+3$ & $\leq 2(n-1)$ \\
    \hline
    Expected latency & $\leq (\frac{3n}{4}+\frac{1}{4})\frac{2^n}{b}$  & $\leq 6\frac{2^n}{b}$ & $\leq 3.585\frac{2^n}{b}$\\
    \hline
  \end{tabular}
  \caption{The comparison of different construction, here $\frac{2^n}{b}<2$. }\label{table1}
  \vspace{-0.1cm}
\end{table*}

Let's construct such a network in the following way: Starting
from a network with single splitter, and at each step adding one more splitter. Assume the current distribution is $\{p_x,p_y,p_z\}$ with $p_x\geq p_y\geq p_z$ (if this is not true, we can change the order of the outputs), then we can add an additional splitter to $p_x$ as shown in Fig. \ref{fig_unBound1}. Iteratively, with $n$ splitters, we can construct a network
with distribution $\{\frac{x}{2^n}, \frac{y}{2^n}, \frac{z}{2^n}\}$ for some integers $x,y,z$ and its expected latency is more than $\frac{n}{3}+\frac{2}{3}$.

By connecting
one output with probability smaller than $\frac{1}{2}$ to the starting point, we can get such a network. \hfill\QED

The theorems above show that the upper bound of the expected latency of a stochastic flow network based on the optimal-sized construction is not well-bounded. However,
this upper bound only reflects the worst case. That does not mean that the optimal-sized construction always has a bad performance in
expected latency when the network size is large. Let's consider the case that the target probability is $\frac{a}{b}$ with $b=2^n$ for some $n$.
In this case, the optimal-sized construction leads to a tree structure, whose expected latency can be written as
\begin{eqnarray*}
ET&=&\sum_{i=1}^{n} \frac{i}{2^i}+\frac{n}{2^n}\\
&=& [\sum_{i=1}^n x^{i+1}]'-\sum_{i=1}^{n-1}\frac{i}{2^i}\\
&=& [\frac{x^2-x^{n+2}}{1-x}]'-\frac{x-x^n}{1-x}\\
&=& 2-\frac{1}{2^{n-1}},
\end{eqnarray*}
which is well-bounded by $2$.

\section{Alternative Constructions}
\label{section_upg}

In the last section, we show that the expected latency of a stochastic flow network based on the optimal-sized construction is not always well-bounded. In this section, we give two other constructions, called
size-relaxed construction and latency-oriented construction. They take both the network size and the expected latency in consideration.
Table \ref{table1} shows the summary of the results in this section, from which we can see that there is a tradeoff between
the upper-bound on the network size and the upper-bound on the expected latency.

\subsection{Size-Relaxed  Construction}

Assume that the desired probability is $\frac{a}{b}$ with $2^{n-1}<b\leq 2^n$ for some $n$. In this subsection, we give a construction, called
size-relaxed construction for realizing $\frac{a}{b}$, with at most $n+3$ splitters and
its expected latency is well-bounded by a constant.

Assume $a$ and $b$ are relatively prime, and let $c=b-a$. Then $\frac{a}{2^n}$ and $\frac{c}{2^n}$ can be represented as binary expansions, namely
$$\frac{a}{2^n}=\sum_{i=1}^{n}a_i 2^{-i},$$
$$\frac{c}{2^n}=\frac{b-a}{2^n}=\sum_{i=1}^{n}c_i 2^{-i}.$$

\begin{figure}[!t]
\centering
\includegraphics[width=1.5in]{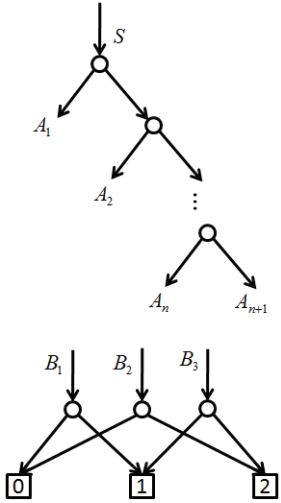}
\caption{The framework to realize probability $\frac{a}{b}$.}
\label{fig_newloop1}
\end{figure}

Let's start from the structure in  Fig. \ref{fig_newloop1}, where the probability of $A_i$ with $1\leq i\leq n$ is $2^{-i}$ and
the probability of $A_{n+1}$ is $2^{-n}$.
We connect $A_i$ with $1\leq i\leq n+1$ to one of \{$B_1,B_2,B_3$ and output $2$\}, such that
the probability distribution of the outputs is $\{\frac{a}{2^{n+1}},\frac{b-a}{2^{n+1}},\frac{2^{n+1}-b}{2^{n+1}}\}$. Based on the values of $a_i,c_i$ with $1\leq i\leq n$ (from binary expansions of $\frac{a}{2^n}$ and $\frac{c}{2^n}$),
we have the following rules for these connections:
\begin{enumerate}
  \item If $a_i=c_i=1$, connect $A_{i}$ with $B_1$.
  \item If $a_i=1, c_i=0$,connect $A_{i}$ with $B_2$.
  \item If $a_i=0, c_i=1$, connect $A_{i}$ with $B_3$.
  \item If $a_i=c_i=0$, connect $A_{i}$ with output $2$.
  \item Connect $A_{n+1}$ with output $2$.
\end{enumerate}

Assume that the probability for a token to reach $B_j$ with $1\leq j\leq 3$ is $P(B_j)$, then we have
$$P(B_1)=\sum_{i=1}^n I_{(a_i=c_i=1)} 2^{-i},$$
$$P(B_2)=\sum_{i=1}^n I_{(a_i=1, c_i=0)} 2^{-i},$$
$$P(B_3)=\sum_{i=1}^n I_{(a_i=0, c_i=1)} 2^{-i},$$
where $I_\phi=1$ if and only if $\phi$ is true, otherwise $I_{\phi}=0$.

As a result, the probability for a token to reach the first output is
$$P_1=\frac{1}{2}(P(B_1)+P(B_2))=\frac{1}{2}\sum_{i=1}^n I_{(a_i=1)}2^{-i}=\frac{a}{2^{n+1}}.$$
Similarly, the probability for a token to reach the second output is
$$P_2=\frac{b-a}{2^{n+1}}.$$

So far, we get that the distribution of the network is $\{\frac{a}{2^{n+1}},\frac{b-a}{2^{n+1}},\frac{2^{n+1}-b}{2^{n+1}}\}$. Similar as Theorem \ref{theorem_gereral1}, by connecting the output $2$ to the starting point, we
get a new network with probability $\frac{a}{b}$. Note that compared to the optimal-sized construction, $3$ more splitters are used in the size-relaxed construction to realize
the desired probability. But it has a much better upper bound on the expected latency as shown in the following theorem.
\begin{Theorem}
Given a network with probability $\frac{a}{b}$ ($2^{n-1}<b<2^n$) constructed using the size-relaxed construction, its expected latency $ET$ is bounded by
$$ET\leq 6\frac{2^n}{b}<12.$$
\end{Theorem}

\proof First, without the feedback, the expected latency for a token to reach $B_1,B_2,B_3$ or output $2$ is less than $2$. This can be obtained from
the example in the last section. As a result, without the feedback, the expected latency for a token to reach one of the outputs is less than $3$. Finally, we can
get the theorem.
\hfill\QED

\begin{figure}[!t]
\centering
\includegraphics[width=2.4in]{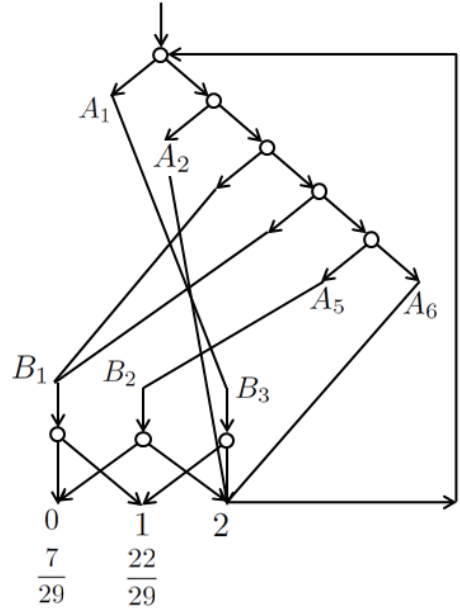}
\caption{The network to realize probability $\frac{7}{29}$.}
\label{fig_upg2}
\end{figure}

Let's give an example of the size-relaxed construction. Assume the desired probability is  $\frac{7}{29}$, then we can write $\frac{a}{2^n}$ and $\frac{b-a}{2^n}$ into binary expansions:
\begin{eqnarray*}
  \frac{a}{2^n}&=&0.00111, \\
  \frac{b-a}{2^n} &=& 0.10110.
\end{eqnarray*}

According to the rules above, we connect $A_1$ to $B_3$, $A_2$ to output $2$,... After connecting output $2$ to the starting point, we can
get a network with probability $\frac{7}{29}$, as shown in Fig. \ref{fig_upg2}.

\begin{figure}[!t]
\centering
\includegraphics[width=1.6in]{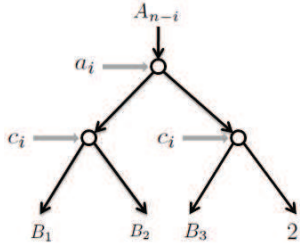}
\caption{The deterministic device to control flow in UPG.}
\label{fig_upg1}
\end{figure}

Another advantage of the size-relaxed construction is that from which we can build  an Universal Probability Generator (UPG) efficiently with $a_i,c_i (1\leq i\leq n)$ as inputs, such that
its probability output is $\frac{a}{a+c}=\frac{a}{b}$. The definition and description of UPG can be found in \cite{Wilhelm2008}. Instead of connecting $A_i$ with $1\leq i\leq n$ to one of $\{B_1,B_2,B_3$ and output $2\}$ directly,
we insert a deterministic device as shown in Fig. \ref{fig_upg1}. At each node of this device, if its corresponding input is $1$, all the incoming tokens will exit the
left outgoing edge. If the input is $0$, all the incoming tokens will exit the right outgoing edge. As a result, the connections between $A_{i}$ and $\{B_1$,$B_2$,$B_3$,Output $2\}$
are automatically controlled by inputs $a_i$ and $c_i$ with $1\leq i\leq n$. Finally, we can get an Universal Probability Generator (UPG), whose output probability is
$$\frac{\sum_{i=1}^{n}a_i 2^{-i}}{\sum_{i=1}^{n}(a_i+c_i)2^{-i}}=\frac{a}{a+c}=\frac{a}{b}.$$

\subsection{Latency-Oriented Construction}

In this subsection, we propose another construction, called latency-orient construction. It uses more splitters than the size-relaxed construction, but achieves
 a better upper bound on the expected latency. Similar to the optimal-sized construction,
this construction is first trying to realize the distribution $\{\frac{a}{2^{n}},\frac{b-a}{2^{n}},\frac{2^{n}-b}{2^{n}}\}$, and then connecting the last output to the starting point.
The difference is that in the latency-oriented construction, this distribution $\{\frac{a}{2^{n}},\frac{b-a}{2^{n}},\frac{2^{n}-b}{2^{n}}\}$ is realized by applying Knuth and Yao's scheme \cite{Knuth1976} that was introduced
in the section of preliminaries.

\begin{figure}[!t]
\centering
\includegraphics[width=2.6in]{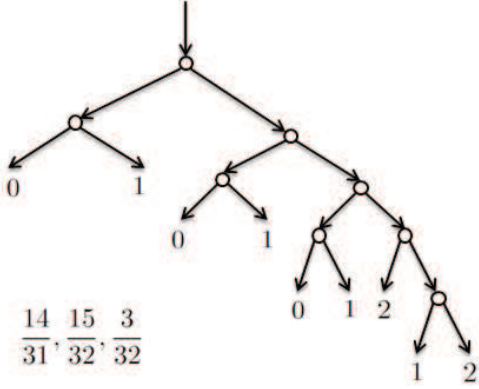}
\caption{The network to realize probability distribution $\{\frac{14}{32},\frac{15}{32},\frac{3}{32}\}$ using Knuth and Yao's scheme.}
\label{fig_schemec}
\end{figure}

Let's go back to the example of realizing probability $\frac{14}{29}$. According to Knuth and Yao's scheme, we need first find the atoms for the
binary expansions of $\frac{14}{32}, \frac{15}{32},\frac{3}{32}$, i.e.
$$\frac{14}{32}\rightarrow (\frac{1}{4},\frac{1}{8},\frac{1}{16}),$$
$$\frac{15}{32}\rightarrow (\frac{1}{4},\frac{1}{8},\frac{1}{16},\frac{1}{32}),$$
$$\frac{3}{32}\rightarrow (\frac{1}{16},\frac{1}{32}).$$

Then we allot these atoms to a binary tree, as shown in Fig. \ref{fig_schemec}. In this tree, the probability for a token to reach outputs labeled $0$ is $\frac{14}{32}$, the probability
for a token to reach outputs labeled $1$ is $\frac{15}{32}$, and the probability
for a token to reach outputs labeled $2$ is $\frac{3}{32}$. If we connect the outputs labeled $2$ to
the starting point, the desired probability $\frac{14}{29}$ can be achieved.

\begin{Theorem}\label{theorem_c}
Given a network with probability $\frac{a}{b}$ ($2^{n-1}<b<2^n$) constructed the latency-oriented construction, its network size is bounded by $2(n-1)
$ and its expected
 latency $ET$ is bounded by
$$ET\leq (log_2 3+2)\frac{2^n}{b}<7.2.$$
\end{Theorem}

\proof Let's first consider the network with distribution $\{\frac{a}{2^n},\frac{b-a}{2^n},\frac{2^n-b}{2^n}\}$, which is constructed using Knuth and Yao's scheme.

1) The network size is bounded by $2(n-1)$. To prove this, let's use $k_j$ to denote
the number of atoms with value $2^{-j}$, and use $a_j$ to denote the number of nodes with depth $j$ in the tree.
Then $k_j$ and $a_j$ have the following recursive relations,
$$a_n=k_n,$$
$$a_j=k_j+\frac{a_{j+1}}{2}, \quad \forall 1\leq j\leq n-1.$$

As a result,
$$\sum_{j=1}^n a_j =\sum_{j=1}^n k_j + \sum_{j=1}^{n-1}\frac{a_{j+1}}{2}.$$

From which, we can get the total number of atoms in the tree is
$$N=\sum_{j=1}^n k_j=\sum_{j=1}^n \frac{a_j}{2}+\frac{a_1}{2}.$$

We know that $k_j$ and $a_j$ also satisfy the following constraints,
$$k_j\leq 3, \forall 1\leq j\leq n,$$
$$a_j \mod 2=0, \forall 1\leq j\leq n.$$

From $j=n$ to $j=1$, by induction, we can prove that
$$a_j\leq 4, \forall 1\leq j\leq n.$$
That is because $a_j$ is even, and if $a_{j+1}\leq 4$, then
$$\frac{a_j}{2}\leq \lfloor \frac{k_j+ \frac{a_{j+1}}{2}}{2}\rfloor\leq 2.$$

Since $a_n, a_1\leq 2$, we can get that
$$N\leq \frac{a_n}{2}+a_1+\sum_{j=2}^{n-1}\frac{a_j}{2}\leq 2n-1.$$

To create $N$ atoms, we need $N-1=2(n-1)$ splitters.

2) The expected latency $ET'$ of the network with distribution $\{\frac{a}{2^n},\frac{b-a}{2^n},\frac{2^n-b}{2^n}\}$ is bounded by $ET'\leq (log_2 3+2)$. That is because
the expected latency $ET'$ is equal to the  expected number of fair bits required. According to the result of Knuth and Yao, it is not hard to get this conclusion.

Now we can get a new network by connecting the last output to the starting point. The
size of the network is unchanged and the expected latency of the new network is $ET=ET'\frac{2^n}{b}$. So we can get the results in the theorem.\hfill\QED

\section{Generating Rational Distributions}
\label{section_distributions}

In this section, we want to generalize our results to generate an arbitrary rational probability distribution $\{q_1,q_2,...,q_m\}$ with $m\geq 2$. Two different methods
will be proposed and studied. The first method is based on Knuth and Yao's scheme and it is a direct generalization of the latency-oriented construction. The second method is based on
a construction with a binary-tree structure. At each inner node of the binary tree, one probability is split into two probabilities. As a result, using a binary-tree structure, the probability one can be split into $m$ probabilities (as a distribution) marked on all the $m$ leaves. In the rest of this section, we will discuss and analyze these two methods.
 Since we consider rational probability distributions, we can write   $\{q_1,q_2,...,q_m\}$ as
$\{\frac{a_1}{b},\frac{a_2}{b}, ..., \frac{a_m}{b}\}$ with integers $a_1,a_2,...b$ and $b$ minimized.

\subsection{Based on Knuth and Yao's scheme}

In order to generate distribution $\{\frac{a_1}{b},\frac{a_2}{b}, ..., \frac{a_m}{b}\}$ with $2^{n-1}<b\leq 2^n$ for some $n$,
we can first construct a network with distribution $\{\frac{a_1}{2^n}, \frac{a_2}{2^n}, ..., \frac{a_m}{2^n}, \frac{2^n-b}{2^n}\}$ using Knuth and Yao's scheme. Then by connecting the
last output to the starting point, we can obtain a network with distribution $\{\frac{a_1}{b},\frac{a_2}{b}, ..., \frac{a_m}{b}\}$. In order to study the properties of this
method, we will analyze two extreme cases: (1) $m=b$ and (2) $m \ll b$.

When $m=b$, the target probability distribution can be written as $\{\frac{1}{b},\frac{1}{b}, ..., \frac{1}{b}\}$. For this distribution, we have the following theorem about
the network constructed using the method based on Knuth and Yao's scheme.

\begin{Theorem}
For a distribution $\{\frac{1}{b},\frac{1}{b}, ..., \frac{1}{b}\}$, the method based on Knuth and Yao's scheme can construct a network with
$b+h(b)-1$ splitters. Here, we assume $b=2^n-\sum_{i=0}^{n-1} \gamma_{i}2^i$ and $h(b)=\sum_{i=0}^{n-1}\gamma_{i}$.
\end{Theorem}

\begin{figure}[!t]
\centering
\includegraphics[width=2.2in]{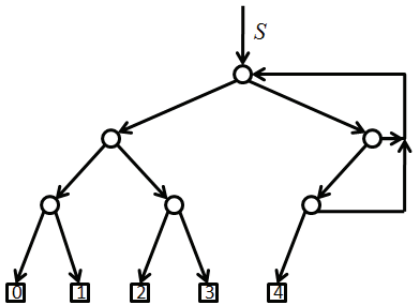}
\caption{The network to realize probability distribution $\{\frac{1}{5},\frac{1}{5},...,\frac{1}{5}\}$.}
\label{fig_example5}
\end{figure}

\proof See the network in Fig. \ref{fig_example5} as an example of the construction.

First, let's consider a complete tree with depth $n$. The network size of such a tree (i.e. the number of parent nodes) is $2^n-1$, denoted by $N_{complete}$.

Let $N(b)$ be the network size of the construction above to realize distribution $\{\frac{1}{b},\frac{1}{b},...,\frac{1}{b}\}$. Assume
$$2^n-b=2^{a_1}+2^{a_2}+...+2^{a_H},$$
with $n>a_1>a_2>...>a_H$ is a binary expansion of $2^n-b$, then we can get
the difference between the size of the construction and the size of the complete binary tree
$$\Delta=N_{complete}-N(b)=\sum_{i=1}^{H}(2^{a_i}-1)=2^n-b-H.$$

So the network size of the construction $N(b)$ is
$$N(b)=2^n-1-(2^n-b-H)=b+H-1,$$
where $H=\sum_{i=0}^{n-1}\gamma_{i}=h(b)$.  \hfill\QED

Let $N^*(b)$ be the optimal size of a network that realizes the distribution $\{\frac{1}{b},\frac{1}{b},...,\frac{1}{b}\}$. It is easy to see that
$N^*(b)\geq b-1$. Note that $h(b)$ is at most the number of bits in the binary expansion of $2^n-b$ (which is smaller than $b$), so we can get the following
inequality quickly
$$b-1\leq N^*(b)\leq N(b)\leq b-1+\log_2 b.$$
It shows that the construction based on Knuth and Yao's scheme is near-optimal when $m=b$. More generally, we believe that when $m$ is
large, this construction has a good performance in network size.

For a general $m$, we have the following results regarding to the network size and expected latency.

\begin{Theorem}
For a distribution $\{\frac{a_1}{b},\frac{a_2}{b}, ..., \frac{a_m}{b}\}$ with $b\leq 2^n$, the method based on Knuth and Yao's scheme can construct a network with
at most $m(n-\lfloor\log_2 m\rfloor+1)$ splitters, such that its expected latency $ET$ is bounded by
$$H(X')\frac{2^n}{b}\leq ET\leq [H(X')+2]\frac{2^n}{b},$$
where $\frac{2^n}{b}<2$. $H(X')$ is the entropy of the distribution $\{\frac{a_1}{2^n},\frac{a_2}{2^n},...,\frac{a_m}{2^n},\frac{2^n-b}{2^n}\}$.
\end{Theorem}

\proof We can use the same argument as that in Theorem \ref{theorem_c}. The proof for the
expected latency is straightforward. Here, we only briefly describe the proof for the network size.

In the network that realizes $\{\frac{a_1}{2^n},\frac{a_2}{2^n},...,\frac{a_m}{2^n},\frac{2^n-b}{2^n}\}$, let's use $k_j$ to denote
the number of atoms with value $2^{-j}$, and use $a_j$ to denote the number of nodes with depth $j$ in the tree.
It can be proved that the total number of atoms in the tree is
$$N=\sum_{j=1}^n k_j=\sum_{j=1}^n \frac{a_j}{2}+\frac{a_1}{2}.$$

Here, the constrains are
$$k_j\leq m+1, \forall 1\leq j\leq n,$$
$$a_j \textrm{ is even}, \forall 1\leq j\leq n.$$

Recursively, we can get that for all $1\leq j\leq n-1$, $a_j\leq 2m$.

For the first $\lfloor\log_2 2m \rfloor$ levels, we have
$$\sum_{j=1}^{\lfloor\log_2 2m \rfloor}a_j\leq 4m.$$

Hence,
\begin{eqnarray*}
N&\leq& \frac{\sum_{j=1}^{\lfloor\log_2 2m \rfloor}a_j}{2}+\frac{a_1}{2}+\frac{\sum_{j=\lfloor\log_2 2m \rfloor+1}^{n}a_j}{2}\\
&\leq& 2m+1+m(n-\lfloor\log_2 2m \rfloor)\\
&\leq& m(n-\lfloor\log_2 m \rfloor+1)+1.
\end{eqnarray*}

So we can conclude that $m(n-\lfloor\log_2 m \rfloor +1)$ splitters are enough for realizing $\{\frac{a_1}{2^n},\frac{a_2}{2^n},...,\frac{a_m}{2^n},\frac{2^n-b}{2^n}\}$
as well as $\{\frac{a_1}{b},\frac{a_2}{b}, ..., \frac{a_m}{b}\}$.
\hfill\QED

This theorem is a simple generalization of the results in Theorem \ref{theorem_c}. Here, the upper bound for the network size is tight only for small $m$.

\subsection{Based on binary-tree structure}

In this subsection, we propose another method to generate an arbitrary rational distribution $\{\frac{a_1}{b}, \frac{a_2}{b},..., \frac{a_m}{b}\}$.
The idea of this method is based on binary-tree structure. We can describe the method in the following way: We construct a binary tree with $m$ leaves, where
the weight of the $i$th $(1\leq i\leq m)$ leaf is $q_i=\frac{a_i}{b}$. For each parent (inner) node, its weight
is sum of the weights of its two children. Recursively,
we can get all the weights of the inner nodes in the tree and the weight of the root node is $1$. For each parent node, assume the weights of its two children are
$w_1$ and $w_2$, then we can replace this parent node by a subnetwork which implements a splitter
with probability distribution  $\{\frac{w_1}{w_1+w_2},\frac{w_2}{w_1+w_2}\}$.
For each leaf, we treat it as an output. In this new network, a token will reach
the $i^{th}$ output with probability $q_i$.

\begin{figure}[!t]
\centerline{\subfigure[]{\includegraphics[width=1.2in]{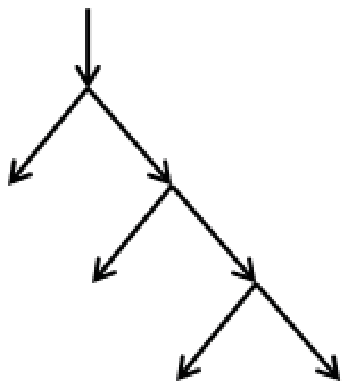} \label{fig_distribution5}}
\hfil \subfigure[]{\includegraphics[width=1.2in]{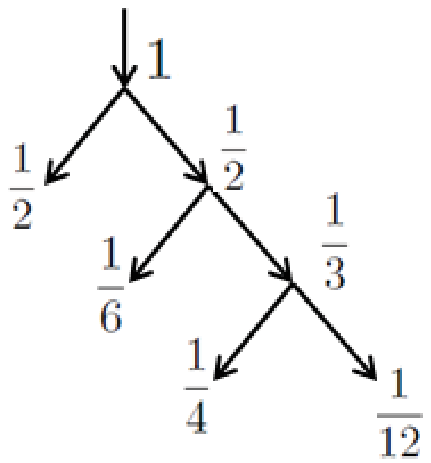} \label{fig_distribution6}}}
\centerline{\subfigure[]{\includegraphics[width=1.3in]{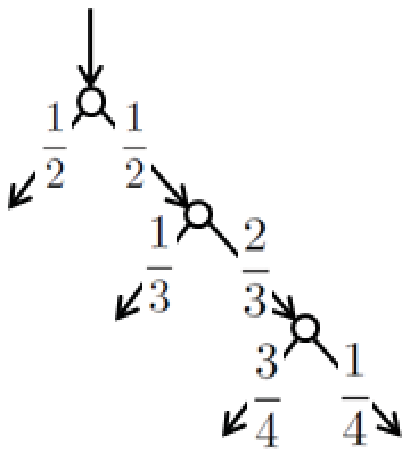} \label{fig_distribution7}}}
\caption{(a) A binary-tree with $4$ leaves. (b) Node weights in the binary tree. (c) The network to realize probability distribution $\{\frac{1}{2},\frac{1}{6},\frac{1}{4},\frac{1}{12}\}$, where $\{\frac{1}{3},\frac{2}{3}\},\{\frac{3}{4},\frac{1}{4}\}$ can be realized using the methods in the sections above. } \label{fig_distribution}
\end{figure}

For example, in order to realize the distribution $\{\frac{1}{2},\frac{1}{6},\frac{1}{4},\frac{1}{12}\}$, we can first generate
a binary-tree with $4$ leaves, as shown in Fig. \ref{fig_distribution5}. Then according to the method above, we can obtain the weight of each node in this binary tree, see Fig. \ref{fig_distribution6}.
Based on these weights, we replace the three parent nodes with three subnetworks, whose probability distributions are $\{\frac{1}{2},\frac{1}{2}\}, \{\frac{1}{3},\frac{1}{3}\}, \{\frac{3}{4},\frac{1}{4}\}$. Eventually, we construct a network with the desired distribution as shown in Fig.  \ref{fig_distribution7}. It can be implemented with $1+2+2=5$ splitters.

\begin{table*}
  \centering
  \renewcommand{\arraystretch}{2}
  \begin{tabular}{|c|c|c|}
    \hline
     & Based on  Knuth and Yao's Scheme & Based on binary-tree structure\\
     \hline
    Network size     & $\leq m(n-\lfloor \log_2 m \rfloor+1) $ & $\leq(m-1)n$  \\
    \hline
    Expected latency & $ \leq(\log_2 (m+1)+2)\frac{2^n}{b}$  & $ \leq (\log_2 m+1)ET_{max} $ \\
    \hline
  \end{tabular}
  \caption{The comparison of different methods, here $\frac{2^n}{b}<2$}\label{table2}
  \vspace{-0.1cm}
\end{table*}

In the procedure above, any binary-tree with $m$ leaves works. Among
all these binary-trees, we need to find one such that the resulting network satisfies our requirements in network size and expected latency.
For example, given the target distribution $\{\frac{1}{2},\frac{1}{6},\frac{1}{4},\frac{1}{12}\}$, the
binary tree depicted above does not result in an optimal-sized construction.
When $m$ is extremely small, such as $3,4$, we can search all the binary-trees with $m$ leaves. However, when $m$ is a little larger, such as $10$, the number of such binary-trees grows exponentially. In this case, the method of brute-force search becomes impractical.
 In the
rest of this section, we will show that Huffman procedure can create a binary-tree with good performances in network size and expected latency for most of the cases.

Huffman procedure can be described as follows \cite{Cover2006}:
\begin{enumerate}
  \item Draw $m$ nodes with weights $q_1,q_2,...,q_m$.
  \item
    Let $S$ denote the set of nodes without parents. Assume node
  $A$ and node $B$ are the two nodes with the minimal weights in $S$,
  then we added a new node as the parent of $A$ and $B$, with weight $w(A)+w(B)$,
  where $w(X)$ is the weight of node $X$.
  \item Repeat 2) until the size of $S$ is $1$.
\end{enumerate}

\begin{figure}[!t]
\centering
\includegraphics[width=2.6in]{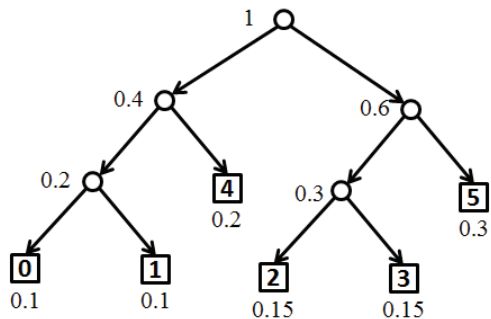}
\caption{The tree constructed using Huffman procedure when the desired distribution is $\{0.1,0.1,0.15,0.15,0.2,0.3\}$}.
\label{fig_distribution3}
\end{figure}

Fig. \ref{fig_distribution3} shows an example
of a binary-tree constructed by Huffman procedure, when the desired distribution is $\{0.1,0.1,0.15,0.15,0.2,0.3\}$.
From \cite{Cover2006}, we know that using Huffman procedure, we can create a tree with minimal expected path length.
Let $EL^*$ denote this minimal expected path length, then its satisfies the following inequality,
$$H(X)\leq EL^*\leq H(X)+1,$$
where $H(X)$ is the entropy of the desired probability distribution $\{q_1,q_2,...,q_m\}=\{\frac{a_1}{b},\frac{a_2}{b},...,\frac{a_m}{b}\}$.

Let $w_i$ denote the weight of the $i^{th}$ parent node in the binary tree.
In order to simplify our analysis, we assume that this parent node can be replaced by a subnetwork with about $\log_2 (bw_i)$ splitters.
This simplification is reasonable from the statistical perspective and according to the  results about our constructions for realizing rational probabilities in the sections above.
Then the size of the resulting network is approximately $\sum_{i=1}^{m-1} \log_2(bw_i)$.
According to Lemma \ref{Lemma_appendix2} in the Appendix, when $m$ is small, Huffman procedure
can create a binary-tree that minimizes $\sum_{i=1}^{m-1} \log_2 w_i$. As a result, among all the binary-trees with $m$ leaves,
the one constructed based on Huffman procedure has an optimal network size -- however, it is only true based on our assumption.
For example, let's consider a desired distribution $\{q_1,q_2,...,q_m\}$ with $\sum_{i\in S}q_i=\frac{1}{2}$ for some set $S$. In this case, the binary-tree structure based on Huffman procedure may not be the best one.

Now we can get the following conclusion about stochastic flow networks constructed using the method based on binary-tree structures.
\begin{Theorem}
For a distribution $\{\frac{a_1}{b},\frac{a_2}{b}, ..., \frac{a_m}{b}\}$ with $b\leq 2^n$, the method based on binary-tree structures constructs a network with
at most $(m-1)n$ splitters. If the binary tree is constructed using  Huffman procedure,
then the expected latency of the resulting network, namely $ET$, is upper bounded by
 $$ET\leq (H(X)+1)ET_{\max},$$
where $H(X)$ is the entropy of the target distribution and $ET_{\max}$ is the maximum expected latency of the inner nodes in the binary-tree.
\end{Theorem}

\proof
1) According to the optimal-sized construction, each inner node can be implemented using at most $n$ splitters.

2) The upper bound on the expected latency is immediate following the result that the expected path length
$EL^*\leq H(X)+1$.
\hfill\QED

\subsection{Comparison}

Let's have a brief comparison between the method based on Knuth and Yao's scheme and the method based on binary-tree structure. Generally, when $m$ is large, the method based Knuth and Yao's scheme may perform better.
When $m$ is small, the comparison between these two methods is given in Table \ref{table2}, where the desired distribution is $\{\frac{a_1}{b},\frac{a_2}{b},...,\frac{a_m}{b}\}$ with $2^{n-1}<b\leq 2^n$. In this table, we assume that the binary tree (in the second method) is constructed using Huffman procedure. $ET_{max}$ denotes the maximum expected
latency of the parent nodes in a given binary-tree.  It is still hard to say that one of the two methods has an absolutely better performance than the other one, no matter in network size or expected latency. In fact, the performance of a construction is usually related with the number structure of the target distribution.
In practice, we can compare both of the constructions based on real values and choose the better one.

\section{Concluding Remarks}
\label{section_conclusion}

Motivated by computing based on chemical reaction networks,
we introduced the concept of stochastic flow networks and
studied the synthesis of optimal-sized networks for realizing
rational probabilities. We also studied the expected latency of
stochastic flow networks, namely, the expected number of splitters a token need to pass before
reaching the output. Two constructions with well-bounded expected latency are proposed. Finally, we generalize our constructions to realize arbitrary
rational probability distributions. Beside of network size and expected latency, robustness is also an important issue in stochastic flow networks. Assume the probability error of each splitter
is bounded by a constant $\epsilon$, the robustness of a given network can be measured by the total probability error. It can be shown that most constructions in this paper are robust
against small errors in the splitters.

To end this paper, we compare a few types of stochastic systems of the same size $n$ in Table \ref{table3}. Here we assume that
the basic probabilistic elements in these systems have probability $1/2$ and we want use them to synthesize the other probabilities. To unfairly compare different systems, we remove
threshold logic circuits from the list, since their complexity is difficult to analyze. From this table, we see that stochastic flow
networks have excellent performances in both expressibility and operating time.
Future works include the synthesis of stochastic flow network to `approximate' desired probabilities or distributions, and the study of the scenario that
the probability of each splitter is not $\frac{1}{2}$.

\begin{table*}
  \centering
  \renewcommand{\arraystretch}{2}
  \begin{tabular}{|c|c|c|}
    \hline
         & Expressibility (probabilities) & Operating time\\
     \hline
    Sequential State Machine \cite{Gill62} & Converge to rational probability $\frac{a}{b}$ with $b\leq n$  & number of states traveled $O(n)$ \\
    \hline
    Stochastic Switching Circuit \cite{Wilhelm2008} & Realize binary probability $\frac{a}{2^n}$ & longest path $O(n)$\\
    \hline
    Combinational Logic \cite{Qian2011} & Realize binary probability $\frac{a}{2^n}$ & maximum depth $O(n)$\\
    \hline
    Stochastic Flow Network & Realize rational probability $\frac{a}{b}$ with $b\leq 2^n$ & expected latency $O(1)$\\
    \hline
  \end{tabular}
  \caption{The comparison of different stochastic systems of size $n$.}\label{table3}
  \vspace{-0.1cm}
\end{table*}

\appendix

\begin{Lemma}
Given $Q$ an $n\times n$ matrix with each entry in $\{0,\frac{1}{2},1\}$, such that sum of each row is at most $1$,
then we have $0\leq det(I-Q) \leq 1$, where $I$ is an identity matrix and $det(\cdot)$ is the determinant of a matrix.
\label{Lemma_appendix1}
\end{Lemma}

\proof
Before proving this lemma, we can see that for any given matrix $Q$, it has the following properties:
For any $i, j$ such that $1\leq i<j\leq n$, switching the $i^{th}$ row with the $j^{th}$ row then switching
the $i^{th}$ column with the $j^{th}$ column, the determinant of $K=I-Q$ keeps unchanged. And more, each entry of $Q$
is still from $\{0,\frac{1}{2},1\}$ and sum of each row of $Q$ is at most $1$. Now, we call the transform above as equivalent
transform of $Q$.

Let's prove this lemma by induction. When $n=1$, we have that
$$Q=\left(
      \begin{array}{c}
        0 \\
      \end{array}
    \right) \textrm{ or } Q=\left(
      \begin{array}{c}
        \frac{1}{2}\\
      \end{array}
    \right) \textrm{ or } Q=\left(
      \begin{array}{c}
        1\\
      \end{array}
    \right).$$
In all of the cases, we have $0\leq det(I-Q)\leq 1$.

Assume the result of the lemma hold for $(n-1)\times (n-1)$ matrix, we want to prove that this result also holds for
$n\times n$ matrix. Now, given a $n\times n$ matrix $Q$, according to the definition in the lemma, we know that the
sum of all the entries in $Q$ is at most $n$. As a result, there exists a column such that the sum of the entries in the column
is at most $1$. Using equivalent transform, we have that
\begin{itemize}
  \item The sum of the entries in the $1^{st}$ column of $Q$ is at most $1$.
  \item The sum of the entries in each row of $Q$ is at most $1$.
\end{itemize}

Now, for the $1^{st}$ column of $I-Q$, let's continue using the equivalent transform to move all
the non-zero entries to the beginning of this column. The possible non-zero entry set of the $1^{st}$ column of $I-Q$
is
$$\phi, \{\frac{1}{2}\},\{1\},\{\frac{1}{2},-\frac{1}{2}\},\{1,-\frac{1}{2}\},\{1,-1\},\{1,-\frac{1}{2},-\frac{1}{2}\}.$$

The first three cases, the result in the lemma can be easily proved. In the following proof, we only consider the other cases
(let $C_1$ denote the non-zero entry set for the $1^{st}$ column of $I-Q$) :

(1) $C_1=\{\frac{1}{2},-\frac{1}{2}\}$.

In this case, we can write $Q$ as
$$Q=\left(
      \begin{array}{cc}
        \frac{1}{2} & A \\
        \frac{1}{2} & B \\
        O & C \\
      \end{array}
    \right)
$$
where $A$ has at most one non-zero entry $-\frac{1}{2}$, the same as $B$.

Let $$E_1=\left(
            \begin{array}{ccccc}
              1 & 0 & 0& \ldots & 0 \\
            \end{array}
          \right)$$
$$I_1=\left(
            \begin{array}{ccccc}
              0 & 1 &0 & \ldots & 0 \\
              0 & 0 &1 & \ldots & 0 \\
              \vdots & \vdots &\vdots & \ddots & \vdots \\
              0 & 0 &0 & \ldots & 1 \\
            \end{array}
          \right)$$

then we have
\begin{eqnarray*}
&&det(I-Q)\\
&=& \frac{1}{2}\det\left(
                         \begin{array}{c}
                         -A\\
                         I_1-C\\
                         \end{array}
                       \right)+ \frac{1}{2}\det\left(
                         \begin{array}{c}
                         E_1-B\\
                         I_1-C\\
                         \end{array}
                       \right)\\
&=& \frac{1}{2}\det\left(
                         \begin{array}{c}
                         E_1-A-B\\
                         I_1-C\\
                         \end{array}\right)\\
&=& \frac{1}{2} \det (I-\left(
                         \begin{array}{c}
                         A+B\\
                         C\\
                         \end{array}\right))
\end{eqnarray*}

Let $D=A+B$, since both $A$ and $B$ has at most one non-zero entry $\frac{1}{2}$, we know that
each entry of $D$ is from $\{0,\frac{1}{2},1\}$, and the sum of all the entries is at most one.
According to our assumption, we know that
$$0\leq \det (I-\left(
                         \begin{array}{c}
                         D\\
                         C\\
                         \end{array}\right)\leq 1.$$
As a result, we have
$$0\leq \det (I-Q)\leq \frac{1}{2}.$$

(2) $C_1=\{1, -\frac{1}{2}\}$.

In this case, we can write $Q$ as
$$Q=\left(
      \begin{array}{cc}
        0 & A \\
        \frac{1}{2} & B \\
        O & C \\
      \end{array}
    \right)
$$

Then
\begin{eqnarray*}
&&det(I-Q)\\
&=& \frac{1}{2}\det\left(
                         \begin{array}{c}
                         -A\\
                         I_1-C\\
                         \end{array}
                       \right)+ \det\left(
                         \begin{array}{c}
                         E_1-B\\
                         I_1-C\\
                         \end{array}
                       \right)\\
&=& \frac{1}{2}\det\left(
                         \begin{array}{c}
                         2E_1-A-2B\\
                         I_1-C\\
                         \end{array}\right)\\
&=& \frac{1}{2}\det(I-\left(
                         \begin{array}{c}
                        A\\
                        C\\
                         \end{array}\right))
+\frac{1}{2}\det(I-\left(
                         \begin{array}{c}
                         2B\\
                         C\\
                         \end{array}\right)).
\end{eqnarray*}

According to our assumption
$$0\leq \det(I-\left(
                         \begin{array}{c}
                        A\\
                        C\\
                         \end{array}\right))\leq 1,$$
$$0\leq \det(I-\left(
                         \begin{array}{c}
                         2B\\
                         C\\
                         \end{array}\right))\leq 1,$$
so $det(I-Q)$ is also bounded by $0$ and $1$.

(3) $C_1=\{1,-1\}$.

Using the same argument as case (1), we can get the result in the lemma.

(4) $C_1=\{1,-\frac{1}{2},-\frac{1}{2}\}$.

In this case, we can write $Q$ as
$$Q=\left(
      \begin{array}{cc}
        0 & A \\
        \frac{1}{2} & B \\
        \frac{1}{2} & C \\
        O & D\\
      \end{array}
    \right)
$$

Let
$$E_2=\left(
            \begin{array}{ccccc}
              0 & 1 & 0& \ldots & 0 \\
            \end{array}
          \right)$$
$$I_2=\left(
            \begin{array}{cccccc}
              0&0 & 1 &0 & \ldots & 0 \\
              0&0 & 0 &1 & \ldots & 0 \\
              \vdots & \vdots & \vdots &\vdots & \ddots & \vdots \\
             0& 0 & 0 &0 & \ldots & 1 \\
            \end{array}
          \right)$$

Then
\begin{eqnarray*}
I-Q=\left(
      \begin{array}{cc}
        1 & -A \\
        -\frac{1}{2} & E_1-B \\
        -\frac{1}{2} & E_2-C \\
        O & I_2-D\\
      \end{array}
    \right)
\end{eqnarray*}

\begin{eqnarray*}
&&det(I-Q)\\&=& \det\left(
                         \begin{array}{c}
                         E_1-B\\
                         E_2-C\\
                         I_2-D\\
                         \end{array}
                       \right)+
                       \frac{1}{2}\det\left(
                         \begin{array}{c}
                         -A\\
                        E_2-C\\
                         I_2-D\\
                         \end{array}
                       \right)\\
&&-
                       \frac{1}{2}\det\left(
                         \begin{array}{c}
                         -A\\
                        E_1-B\\
                         I_2-D\\
                         \end{array}
                       \right) \\
&=& \frac{1}{2} \det\left(
                         \begin{array}{c}
                         E_1-B-A\\
                         E_2-C\\
                         I_2-D\\
                         \end{array}
                       \right)+
                       \frac{1}{2}\det\left(
                         \begin{array}{c}
                         E_1-B\\
                        E_2-C-A\\
                         I_2-D\\
                         \end{array}
                       \right)\\
\end{eqnarray*}

Now, we can write $A=E+F$ such that both $E$ and $F$ has at most one non-zero entry, which is
$\frac{1}{2}$.  Therefore,

\begin{eqnarray*}
&&det(I-Q)\\
&=& \frac{1}{2} \det\left(
                         \begin{array}{c}
                         E_1-B-E-F\\
                         E_2-C\\
                         I_2-D\\
                         \end{array}
                       \right)\\
&&+
                       \frac{1}{2}\det\left(
                         \begin{array}{c}
                         E_1-B\\
                        E_2-C-E-F\\
                         I_2-D\\
                         \end{array}
                       \right)
\end{eqnarray*}
where
\begin{eqnarray*}
&&\det\left(
                         \begin{array}{c}
                         E_1-B-E-F\\
                         E_2-C\\
                         I_2-D\\
                         \end{array}
                       \right)\\
&=& \det\left(
                         \begin{array}{c}
                         E_1-B-E\\
                         E_2-C-F\\
                         I_2-D\\
                         \end{array}
                       \right)+ \det\left(
                         \begin{array}{c}
                         -F\\
                         E_2-C\\
                         I_2-D\\
                         \end{array}
                       \right)\\
& & +  \det\left(
                         \begin{array}{c}
                         E_1-B-E\\
                         F\\
                         I_2-D\\
                         \end{array}
                       \right)
\end{eqnarray*}
and
\begin{eqnarray*}
&&\det\left(
                         \begin{array}{c}
                         E_1-B\\
                        E_2-C-E-F\\
                         I_2-D\\
                         \end{array}
                       \right)     \\
&=&\det\left(
                         \begin{array}{c}
                         E_1-B-F\\
                         E_2-C-E\\
                         I_2-D\\
                         \end{array}
                       \right)+\det\left(
                         \begin{array}{c}
                         E_1-B\\
                         -F\\
                         I_2-D\\
                         \end{array}
                       \right)\\
& & + \det\left(
                         \begin{array}{c}
                         F\\
                         E_2-C-E\\
                         I_2-D\\
                         \end{array}
                       \right)
\end{eqnarray*}
Finally, we can get that
\begin{eqnarray*}
&&det(I-Q)\\
&=& \frac{1}{2} \det[I-\left(
                         \begin{array}{c}
                         B+E\\
                         C+F\\
                         D\\
                         \end{array}
                       \right)]+\frac{1}{2} \det[I-\left(
                         \begin{array}{c}
                         B+F\\
                         C+E\\
                         D\\
                         \end{array}
                       \right)]
\end{eqnarray*}

According to our assumption, we have that
$$0\leq \det[I-\left(
                         \begin{array}{c}
                         B+E\\
                         C+F\\
                         D\\
                         \end{array}
                       \right)]\leq 1,$$
$$0\leq \det[I-\left(
                         \begin{array}{c}
                         B+F\\
                         C+E\\
                         D\\
                         \end{array}
                       \right)]\leq 1.$$
Therefore, the result of this lemma holds.

This completes the proof. \hfill\QED

\begin{Lemma}
Given a desired probability distribution $\{q_1,q_2,...,q_m\}$ and $m<6$, Huffman procedure can construct a binary-tree such that
\begin{enumerate}
  \item It has $m$ leaves with weight $q_1,q_2,...,q_m$.
  \item $L=\sum_{j=1}^{m-1}\log_2 w_j$ is minimized, where $w_j$ is the weight of $j^{th}$ parent node in a binary tree with $m$ leaves.
\end{enumerate}
\label{Lemma_appendix2}
\end{Lemma}

\begin{figure}[!h]
\centering
\includegraphics[width=3.2in]{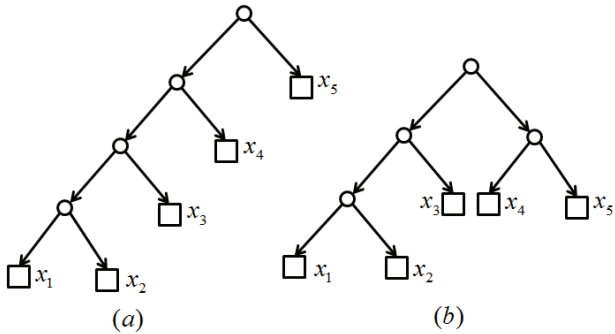}
\caption{Two possible tree structures for $m=5$.}
\label{fig_distribution4}
\end{figure}

\proof It is easy to prove that the case for $m=3$ or $m=4$ is true. In the following proof, we only show the case
for $m=5$ briefly. W.l.o.g, we assume $q_1\leq q_2\leq ...\leq q_5$. Without considering
the order of the leaves, we have only two binary-tree structures, as shown in Fig. \ref{fig_distribution4}.

In both of the structures, for any pair of leaves $x_i$ and $x_j$, if $x_i$'s sibling is $x_j$'s ancestor then
$x_i\geq x_j$. Otherwise, we can switch the position of $x_i$ and $x_j$ to reduce $\sum_{j=1}^{m-1}\log_2 w_j$.
So if the tree structure (a) in Fig. \ref{fig_distribution4} is the optimal one, we have $x_1=q_1, x_2=q_2$ or $x_1=q_2, x_2=q_1$. Now, we will show
that if the tree structure (b) in Fig. \ref{fig_distribution4} is the optimal one, we also have $x_1=q_1, x_2=q_2$ or $x_1=q_2, x_2=q_1$.

For the tree structure (b), we have the following relations:
$$x_3\geq \max\{x_1,x_2\},$$
$$x_4+x_5\geq \max\{x_1+x_2,x_3\}.$$
Then $q_1$ and $q_2$ is in $\{x_1,x_2,x_4,x_5\}$ and $x_1+x_2\leq \frac{1-x_3}{2}$.

Let $x=x_1+x_2$, then $L$ can be written as
\begin{eqnarray*}
L&=&\min \log (x_1+x_2)+\log (x_1+x_2+x_3)+\log (x_4+x_5)\\
&=& \min \log ((x_1+x_2)(x_1+x_2+x_3)(1-x_1-x_2-x_3))\\
&=& \min \log x(1-x_3-x)(x+x_3).
\end{eqnarray*}

So we can minimize
$x(1-x_3-x)(x+x_3)$ instead of minimizing $L$. Fixing $x_3$, we can see that
$x(1-x_3-x)$ increases as $x$ increases when $x\leq \frac{1-x_3}{2}$;
$(x+x_3)$ also increases as $x$ increases. So fixing $x_3$, $x(1-x_3-x)(x+x_3)$
is minimized if and only if $x$ is minimized, which will cause $x_1=q_1, x_2=q_2$
or $x_1=q_2,x_2=q_1$.

Based on the discussion above, we know that in the optimal tree, $q_1$ and $q_2$ must be
siblings. Let's replace $q_1$, $q_2$ and their parent node using a leaf with weight $q_1+q_2$. Then we can get an optimal tree
for distribution $\{q_1+q_2,q_3,q_4,q_5\}$, whose $L$ value is $L_4^*$. Assume the optimal $L$ value
for distribution $\{q_1,q_2,q_3,q_4,q_5\}$ is $L_5^*$, then
$$L_5^*=L_4^*+\log_2(q_1+q_2).$$

Let's consider a tree constructed by Huffman procedure for $\{q_1,q_2,q_3,q_4,q_5\}$, whose $L$ value is $L_5$. We want to show that this tree is optimal. According to the
procedure, we know that $q_1$ and $q_2$ are also siblings. By combing $q_1$ and $q_2$ to a leaf with $q_1+q_2$,
we can get a new tree. This new tree can be constructed by applying Huffman procedure to distribution $\{q_1+q_2,q_3,q_4,q_5\}$.
Due to our assumption for $m=4$, it is optimal, as a result the following result is true,
$$L_5=L_4^*+\log_2(q_1+q_2).$$

Finally, we can obtain $L_5=L_5^*$, which shows that the $L$ value of the tree constructed by Huffman procedure is minimized when $m=5$.

This completes the proof.\hfill\QED


\begin{thebibliography}{11}
\bibitem{Abrahams1996}
J. Abrahams. ``Generation of discrete distributions from biased coins", \emph{IEEE Transactions on Information Theory},
vol. 42, pp. 1541-1546, 1996.

\bibitem{Cover2006}
T. M. Cover, J. A. Thomas. ``Elements of information theory", Wiley-Interscience, Second Edition, 2006.

\bibitem{Elias1972}
P. Elias, ``The efficient construction of an unbiased random sequence", \emph{Ann. Math. Statist.}, vol. 43, pp. 865-870, 1972.

\bibitem{Gill62}
A. Gill, ``Synthesis of probability transformers", \emph{Journal of the Franklin
Institute}, vol. 274, no. 1, pp. 1-19, 1962.

\bibitem{Gill63}
A. Gill, ``On a weight distribution problem, with application to the design
of stochastic generators", \emph{Journal of the ACM}, vol. 10, no. 1, pp. 110-121, 1963.

\bibitem{Han1997}
T. S. Han and M. Hoshi, ``Interval algorithm for random number generation", \emph{IEEE Trans. on Information Theory}, vol. 43, No. 2, pp. 599-611, 1997.

\bibitem{Hoeffding1970}
W. Hoeffding and G. Simon, ``Unbiased coin tossing with a biased coin", \emph{Ann. Math. Statist.}, vol. 41, pp. 341-352, 1970.

\bibitem{Knuth1976}
D. Knuth and A. Yao, ``The complexity of nonuniform random number generation", \emph{Algorithms and Complexity: New Directions and Recent Results}, pp. 357-428, 1976.

\bibitem{Loh2009}
P. Loh, H. Zhou, and J. Bruck, ``The robustness of stochastic switching networks," in \emph{Proc. IEEE International Symposium on Information Theory (ISIT)}, 2009.

\bibitem{Neumann1951}
J. von Neumann, ``Various techniques used in connection with random digits", \emph{Appl. Math.
Ser.}, Notes by G. E. Forstyle, Nat. Bur. Stand., vol. 12, pp. 36-38, 1951.

\bibitem{Peres1992}
Y. Peres, ``Iterating von Neumann's procedure for extracting random bits", \emph{Ann. Statist.}, vol. 20, pp. 590-597, 1992.

\bibitem{Qian2011}
W. Qian, M. D. Riedel, H. Zhou, and J. Bruck, ``Transforming probabilities with combinational logic",
\emph{IEEE Trans. on Computer-Aided Design of Integrated Circuits and Systems}, vol. 30, pp. 1279-1292, 2011.

\bibitem{Sheng1965}
C. L. Sheng, ``Threshold logic elements used as a probability transformer",
\emph{Journal of the ACM}, vol. 12, no. 2, pp. 262-276, 1965.

\bibitem{Soloveichik2009}
D. Soloveichik, G. Seelig, and E.Winfree, ``DNA as a universal substrate for chemical kinetics", \emph{LNCS 5347}, pp. 57-69, 2009.

\bibitem{Valkenburg1974}
M. E. Van Valkenburg, ``Network analysis", 3rd Edition, Prentice-Hall, Englewood Cliffs, NJ, USA, 1974.

\bibitem{Warren1984}
Q. Stout and B. Warren, ``Tree algorithms for unbiased coin tosssing with a biased coin", \emph{Ann. Probab.}, vol. 12, pp. 212-222, 1984.

\bibitem{Wilhelm2008}
D. Wilhelm and J. Bruck, ``Stochastic switching circuit
synthesis", in \emph{Proc. IEEE International Symposium on Information Theory (ISIT)},  pp. 1388-1392, 2008.

\bibitem{Zhou09}
H. Zhou, J. Bruck. ``On the expressibility of stochastic switching circuits", in \emph{Proc. IEEE International Symposium on Information Theory (ISIT)}, pp. 2061-2065, 2009.

\bibitem{Zhou2011}
H. Zhou, P. Loh, and J. Bruck, ``The synthesis and analysis of stochastic switching circuits," Technical Report, California Institute of Technology, 2011.
\end{thebibliography}
\end{document}